\documentclass[showpacs,twocolumn,nofootinbib,superscriptaddress]{revtex4}
\usepackage{amsmath,amssymb,amsbsy,graphics,graphicx,epsfig,array,latexsym,booktabs}
\usepackage[colorlinks,hypertex]{hyperref}
\usepackage{ifthen,color,relsize,fancyvrb}
\hypersetup{pdfpagemode=FullScreen,urlcolor=black,citecolor=blue}

\begin{document}
\title{Probing $e^{+}e^{-}$ annihilation in noncommutative electroweak model}

\author{Chien Yu Chen}
\email{d9522817@phys.nthu.edu.tw}
\affiliation{\small Department of Physics, National Tsing-Hua University, Hsinchu 300, Taiwan}
	
\preprint{\hepth{0808.2848}} 

\begin{abstract}
If the twist $Poincar\acute{e}$ transformation is imposed on the
noncommutative spacetime, then $Lorentz$ invariance cannot be
applied on QFT. To data, noncommutative theory is one of the best
candidates to modify $Lorentz$ transformation. In this paper, we
argue parity violation under the process of $e^{+}e^{-}\to\gamma
\gamma$ and make a detailed analysis of the difference behavior
of each helicity state on noncommutative spacetime. The effect
arises from the production of spin and magnetic fields. We check
the energy momentum conservation for all used couplings and
discover that if the electric field changes particle energy
spectrum, there is no symmetry violation as the field produces
a longitudinal state on the finial triple boson couplings.
\end{abstract}

\maketitle

\section{Introduction}
$Lorentz$ symmetry constrains the transformation of spacetime from
boost and rotation. Many phenomena that cannot be predicted by the
standard model are expected to violate $Lorentz$ symmetry. The main
purpose of this paper is to concentrate on the effects
induced by the background magnetic field. In particular, we discuss the
parity asymmetry in $e^{+}e^{-}\to\gamma\gamma$ with the content
of $Lorentz$ violation. The numerical results present that parity is
violated, while $CP$ symmetry still preserves in the next leading
$\theta_{\mu\nu}$ patch with noncommutative background.

Furthermore, $CP$ symmetry puts a constraint on the cross section.
Parity violated phenomena simultaneously violates the charge
conservation. This effect induces a slight space transilation, but
charge violated event changes the magnitude in total cross section. In
the viewpoint of quantum gravity, energy scale of $Lorentz$
violation is ranged in the Plank scale, $M_{PL} = 10^{19}~GeV$. In
this paper, we probe the effects of backgound field direction
on the total cross section. The scale $\Lambda_{C}$ and the colliding
energy level have been set to 1~$TeV$ and 800~$GeV$ respectively.
The total cross section fluctuation associates with
$70K_{Z\gamma\gamma}\cos\alpha_{B}$ ($fb$) in summing each
photon polarization, where $K_{Z\gamma\gamma}$ is a triple gauge
boson coupling, and $\alpha_{B}$ is the direction of background
magnetic field. The shift is miniscule in comparing with the standard
model cross section 5560 ($fb$).

Moreover, due to the interaction between background magnetic field and
photons with oppsite polarization, the spin-magnetic interaction, $\vec{S}
\cdot\vec{B}$, produces a forward-backward asymmetry cannot be
predicted by the standard model. In calculations, the sensitive
phenomenon of the total cross section of central energy is pertaining
to the direction of background magnetic field. We observe that the
spin-magnetic interaction effect is changed under the relation, $<\vec{S}>
\cdot\vec{B}$ = $\pm\arrowvert B\arrowvert\cos\theta_{B_{Z}}$. The
principal frame takes along z-axis, and the total energy spectrum is proportional
to the $\theta_{B_{Z}}$ angle. By the way, the electric field is absent
due to the unitarity constraint.

Most investigations tend to choose a preferred direction of the
isotropic and homogeneous earlier universe, i.e. by adding a nonlocal
four vector term in the Lagrange~\cite{Seiberg:1999vs}. The
field along the direction of the background field equally imposes a
constant direction rearranging the order of spacetime~\cite{Minwalla:1999px}.
There are some papers consider noncommutative scalar field in $fuzzy$
sphere~\cite{Panero:2006bx}, this spacetime considers the era of universe
earlier than cosmology scale. However, many theories with consistent
concepts are to define a preferred direction on the isotropic spacetime.
This is apparently to oppose the general assumption of $Lorentz$ symmetry.
Noncommutative field theory is one of the theories violates $Lorentz$
symmetry in putting a constant background field term in the
$Dirac~Born~Infeld$ action of the bosonic string~\cite{Seiberg:1999vs}.
The field influences the position between particles cannot be
exchanged on the same consequence.

In the concept of noncommutive spacetime, there are three kinds of structure
~\cite{Jurco:2000ja} considered: (1) canonical structure,
(2) Lie algebra structure, and (3) quantum space structure. It dominates
to decide a way by $\star$ production. We cannot think of a different description of the
gauge transformation into a different map. On the model building, unfortunately,
the noncommutative model has been restricted by the $\textit{No-Go Theorem}$
~\cite{Chaichian:2001mu}. Only $U_{\star}(1)$ gauge group can
build into this spacetime. Separating the generator into U(1) gauge and $SU_{\star}(N)$
parts, the redefined relation between each particle can be formed a group to aside
a existence of condensed field. Requiring non-abelian
representation in considering enveloping algebra
~\cite{Jurco:2000ja,Wulkenhaar:1999im}, it separates the
generators of different commutation relations and expands the
gauge representation of infinite $\theta_{\mu\nu}$ deformation.
Renormalization implies unitary constraint is satisfied in the field
theory and restricts $\theta_{\mu\nu}$ is just considered
into first order.

The UV/IR mixing commits the particle to propagate nonlocal space.
The mixed angular momentum from background constant direction and particle kinetic
momentum towards to renew the commutation form or split the $U(1)$ generation.
Due to slip the photon polarization under modifying commutation relation will absorb
some physical degrees of freedom into the lost generators, the complete physical field is
still in the unbroken gauge. Hereafter, the redefined angular momentum is considerable
to modify $U(1)$ gauge. We do not take the condensed picture in gauge boson, the couplings used
concentrate on the unbroken $U(1)$ generators with preserving chiral symmetry in fermion fields.
Hence, the behavior of finial photon polarization is the simple consequence of background nonlocal
vector condensation with particle polarizations.

Expanding the origin nonabelian gauge theory in noncommutative
spacetime and considering the enveloping algebra modifies gauge
representation. Using $Seiberg~Witten$ map~\cite{Seiberg:1999vs, Martin:2002nr},
the origin gauge group of the standard model is extended by first order
$\theta_{\mu\nu}$ deformation under noncommutative phase-like translation,
\begin{equation}
f(x) \star g(x)=f(x)exp\bigg(\frac{i}{2}
\overleftarrow{\partial_{\mu}}\theta^{\mu\nu}\overrightarrow
{\partial_{\mu}}\bigg)g(x).\nonumber
\end{equation}
Which are two function products of $\star$ deformed noncommutative
algebra. The gauge group is extended as $SU_{\star C}$(3)
$\otimes SU_{\star L}$(L)$\otimes U_{\star Y}$(1) with produced
background deformation by preserving gauge restriction. In the next section,
we briefly introduce gauge boson action using
enveloping algebra expansion~\cite{Jurco:2000ja}. Thereof,
all of the field theory involving $\theta_{\mu\nu}$
deformation comments the physics ordered phase, and contains the
information of earlier universe background magnetic and electric field.

Using the properties of space and momentum exchange under Moyal space,
the $Lorentz$ group SO(1,3) is isomorphic to O(1,1)$\otimes$SO(2), where the
lost generators are residing in the hypersurface. It results from an
arbitrary generator and uniquely choose in the background field direction
and violates boost and rotation symmetry. These phenomena induce
parity symmetry violated effects. The common commutation relation
on the four vector spacetime is
\begin{equation}\label{eq1}
[x_{\mu}, x_{\nu}]_{\star} = i\theta_{\mu\nu}=i\frac{C_{\mu\nu}}
{\Lambda^{2}_{NC}},
\end{equation}
and
\begin{equation}\label{eq2}
C_{\mu\nu} = \left(\begin{array}{@{}cccc@{}}
0 & E_{1} & E_{2} & E_{3}\\
-E_{1} & 0 & -B_{3} &B_{2}\\
-E_{2} & B_{3} & 0 & -B_{1}\\
-E_{3} & -B_{2} & B_{1} & 0
\end{array}\right),
\end{equation}
where $\theta_{\mu\nu}$ contains all the information of the background
field, such as the field strength tensor of electrodynamics. The cross
section is charge violated due to parity violation and $CP$
conservation of odd order theta deformation.

However, the spin of any physics field interacts with background field
in odd order $\theta_{\mu\nu}$ deformations. If we choose a preferred
direction on the homogeneous and isotropic spacetime, under the background,
parity does not remain a perfectly symmetry. Each particle helicity induces an
opposite contribution on coupling to the background field. Particle
energy spectrum is exchanged by the spin and background magnetic interaction.
Therefore, if the deviation of each helicity dispersion is the same, the
total parity violated phenomenon will be invisible. On the other hand, 
each photon helicity induces an opposite contribution on forward-backward
asymmetry, the unpolarized electron initial beams will produce an
asymmetric deviation to each helicity of photon luminosity.

\section{Brief review of noncommutative theory}
On the commutative spacetime we use the Seiberg-Witten map to
generate noncommutative theta deformed potential. The series ordered
$\theta_{\mu\nu}$ expansion in enveloping algebra extends
the non-abelian gauge symmetry from $SU(2)\otimes U(1)$
to $SU_{\star L}$(L)$\otimes U_{\star Y}$(1)
~\cite{Jurco:2000ja, Wulkenhaar:1999im}. The standard noncommutative
model is invariant under the gauge transformation builded by Hopf algebra
~\cite{Wulkenhaar:1999im, Martin:2002nr, Zahn:2006wt} on Moyal
space~\cite{Wallet:2007em}. It supposes the existence of an infinitesimal
transformation generator $X$ with $\phi$ $\longmapsto$
X$\vartriangleright$$\phi$. The action of the field is multiplied
by a coproduct $\triangle$, denoted in
$\phi\otimes\psi\longmapsto\triangle$(X)$\vartriangleright$($\phi
\otimes\psi$).

The translation of coproduction between the twist deformation
and the initial form,
\begin{equation}\label{eq3}
\triangle_{\theta}(X) = \mathfrak{F}^{-1}\triangle_{0}(X)\mathfrak{F} =
\mathfrak{F}^{-1}(X\otimes 1+1\otimes X)\mathfrak{F},
\end{equation}
and the noncommutative momentum translation representation,
\begin{equation}\label{eq4}
\mathfrak{F} = exp(-\frac{i}{2}\theta^{ij}p_{i}\otimes p_{j}),
\end{equation}
are defined by abelian gauge transformation. The coproduct of 
Poincar$\acute{e}$ generator requires a consistent deformation 
between two fields, $m_{0}(\phi\otimes\psi)$ = $\phi\cdot\psi$, and
isomorphic to $m_{\theta}(\phi\otimes\psi)$ = $\phi\star\psi$. 
Therefore, the translation of the gauge symmetry under this rule is
similarly to take Eq.(2.1) and Eq.(2.2) into
\begin{equation}\begin{split}\label{eq5}
&X\vartriangleright  m_{0}(\phi\otimes\psi) = m_{0}(\triangle_{0}(X)
\vartriangleright(\phi\otimes\psi ))\nonumber\\
&\longmapsto X\vartriangleright  m_{\theta}(\phi\otimes\psi) =
m_{0}(\triangle_{\theta}(X)\vartriangleright(\phi\otimes\psi )).
\end{split}\end{equation}
We use this representation to prove photon polarization does 
not be changed in the noncommutative spacetime. However, 
if $\psi$ and $\phi$ are substituted for four vector momentum, 
and $Pauli-Ljubanski$ polarization four vectors individually,
\begin{equation}\label{eq6}
\mathbb{W}^{\mu} = \frac{1}{2}\epsilon^{\mu\nu\alpha\beta}
J_{\nu\alpha}P_{\beta},
\end{equation}
in which $J_{\nu\alpha}$ is $Lorentz$ rotation and boosts 
generator, $P_{\beta}$ is the momentum operator.

Hence, after transformation it is easily shown that, for 
chargless particle, $Lorentz$ tensor violates the origin 
translation and rotation in isotropic and homogeneous spacestime. 
On the other hand, if the field contains a charge, momentum 
translation is violated along the background electric field 
direction. Photon is a chargeless particle, the direction of translation 
will not induce another degree of freedom to generate its mass.
In fact that noncommutative is translational invariance in 
Eq.(2.1, 2.2, 2.3). Following above discussion, $m_{\theta}$
($P_{\mu}\otimes P_{\nu}$ - $P_{\nu}\otimes P_{\mu}$) = 0
takes a constraint on $Pauli-Ljubanski$ polarization. The commutation
relations $m_{\theta}$($\mathbb{W}^{\mu}\otimes P^{\nu}$-
$P^{\nu}\otimes\mathbb{W}^{\mu}$) = 0, $P^{2}$ = $m^{2}$ and
$\mathbb{W}^{2}$  = $m^{2}$s(s+1) still retain the properties of
$Casimir$ operator, where m is particle mass along to the direction of
momentum and s is its polarization. For the massless case,
$\mathbb{W}^{2}$  = 0, and m = 0, photon does not contain a
longitudinal state even after momentum translation. Therefore,
gauge condition $m_{\theta}(P^{\mu}\otimes\mathbb{W}_{\mu})$ = 0
is still unchanged. However, the summation of polarization should
add a phase $\phi\thicksim\vec{B}\cdot(\vec{P}_{1}\times\vec{P}_{2})$
due to two gauge bosons product.

The noncommutative gauge theory is very interesting in which
contains many degrees of freedom from choosing a 
different representations of gauge kinetic term under trace technique. 
On this way, we use the enveloping algebra to realize the 
nonabelian group\cite{Jurco:2000ja}, and choose a minimal expression of 
gauge expansion. By dividing the gauge kinetic term, one part is minimal
and another is non-minimal. The gauge action of noncommutative 
electroweak model\cite{Calmet:2001na} is regarded as
\begin{equation}\label{eq7}
S_{gauge} = S^{minimal}_{gauge} + S^{nm-term}_{gauge},
\end{equation}
the minimal term is to expand the origin using 
$Seiberg~Witten$ map. In order to consider a
triplet gauge boson couplings, hence, the non-minimal term is to choose
a different trace technique on the aspect of gauge boson
parameter to expand the gauge boson action,
\begin{equation}\begin{split}\label{eq8}
&S^{minimum}_{gauge} =\\
&- \frac{1}{2}\int d^{4}x\bigg{(}\frac{1}{2} A_{\mu\nu}A^{\mu\nu} +
Tr B_{\mu\nu}B^{\mu\nu} + Tr G_{\mu\nu}G^{\mu\nu}\bigg{)}\\
&+ \frac{1}{4}g_{s}d^{abc}\theta^{\rho\sigma}\int d^{4}x\bigg{(}
\frac{1}{4}G^{a}_{\rho\sigma}G^{b}_{\mu\nu} - G^{a}_{\rho\mu}
G^{b}_{\sigma\nu}\bigg{)}G^{\mu\nu,c} + O(\theta^{2}),\nonumber
\end{split}\end{equation}
and
\begin{equation}\begin{split}\label{eq9}
&S^{nm-term}_{gauge} =\\
&g'^{3}k_{1}\theta^{\rho\sigma}\int d^{4}x\bigg{(}\frac{a}{4}A_{\rho\sigma}
A_{\mu\nu} - A_{\mu\rho}A_{\nu\sigma}\bigg{)}A^{\mu\nu}\\
&+ g'g^{2}k_{2}\theta^{\rho\sigma}\int d^{4}x\bigg{[}\bigg{(}\frac{a}{4}
A_{\rho\sigma}B^{a}_{\mu\nu} - A_{\mu\rho}B^{a}_{\nu\sigma}\bigg{)}
B^{\mu\nu,a} + c.p.\bigg{]}\nonumber\\
&+ g'g_{s}^{2}k_{3}\theta^{\rho\sigma}\int d^{4}x\bigg{[}\bigg{(}
\frac{a}{4}A_{\rho\sigma}G^{b}_{\mu\nu} - A_{\mu\rho}G^{b}_{\nu\sigma}
\bigg{)}G^{\mu\nu,b} + c.p.\bigg{]}\\
&+ O(\theta^{2}),\nonumber\\
\end{split}\end{equation}
the first one is the origin gauge boson kinetic term on 
noncommutative spacetime, the parameter "a" is an extra gauge
degrees of freedom. In this paper, we set the constant parameter
to 3 by imposing renormalization and unitary conditions. Another
is non-minimal term, considering the freedom of different trace technique
on kinetic gauge field to construct the non-minimal version of
$mNCSM$ in using the different Seiberg-Witten map.

Each triple gauge boson coupling is derived from the above action.
Extracting the couplings from the Lagrange, couplings of
$\gamma-\gamma-\gamma$ and $Z-\gamma-\gamma$ are presented
as follows,
\begin{equation}\label{eq10}
\mathfrak{L}_{\gamma\gamma\gamma} = \frac{e}{4}\sin2\theta_{W}
K_{\gamma\gamma\gamma}\theta^{\rho\sigma}A^{\mu\nu}\big{(}
aA_{\mu\nu}A_{\rho\sigma} - 4A_{\mu\rho}A_{\nu\sigma}\big{)}
\end{equation}
\begin{equation}\begin{split}\label{eq11}
\mathfrak{L}_{Z\gamma\gamma} &= \frac{e}{4}\sin2\theta_{W}
K_{Z\gamma\gamma}\theta^{\rho\sigma}\big{[}2Z^{\mu\nu}
\big{(}2A_{\mu\rho}A_{\nu\sigma} - aA_{\mu\nu}A_{\rho\sigma}\big{)}
\nonumber\\
                                               &+ 8Z_{\mu\rho}A^{\mu\nu}A_{\nu\sigma}
                                                - aZ_{\rho\sigma}A^{\mu\nu}A_{\mu\nu}\big{]},
\end{split}\end{equation}
where the couplings $K_{\gamma\gamma\gamma}$ and $K_{Z\gamma\gamma}$ 
contain the gauge parameters, $g$, $g_{s}$, and $g'$.
Ref.\cite{Duplancic:2003hg} plots the range of 
all these couplings and also makes more detailed analysis to give a constraint.
The couplings are composed by $g_{i}$, $i$ goes from 1 to 6. The $C$ and $P$
are violated in these couplings, but preserves $CP$ symmetry in non-planar
tree level diagram.

On the Seiberg-Witten map, there are some kinds of coupling
induced by the connection of $\theta_{\mu\nu}$.
Following the electroweak model~\cite{Calmet:2001na}, the change
up to the first order $\theta_{\mu\nu}$ modification uses the
enveloping algebra to extend the non-abelian gauge group. The
interesting coupling Z-$\gamma$-$\gamma$ violates the angular
momentum distribution\cite{Snyder:1946qz}, hence it is exactly forbidden
on the commutative standard model. Approximately, the branching radio
of Z $\to$ $\gamma\gamma$ is 4$\times 10^{-8}$, and the range of
coupling are -~0.333$<K_{Z\gamma\gamma}<$0.095 and -~0.184$
<K_{\gamma\gamma\gamma}<$-~0.419\cite{Duplancic:2003hg}. In this paper,
we set $K_{Z\gamma\gamma}$ = -~0.2, and $K_{\gamma\gamma\gamma}$
= -~0.3 for convenient.

In renormalization aspects, the triple coupling tensor 
$\Theta^{\mu\nu\rho}$ is changed by choosing a different 
map~\cite{Martin:2002nr, Kulish:2006jq}. However, the map produces
a geometric freedom in gauge sector. The triple gauge boson coupling tensor
is
\begin{equation}\begin{split}
&\Theta^{\mu\nu\rho}_{3}(a; k_{\mu 1},k_{\nu 2},k_{\rho 3}) =\\
&-(k_{1}\theta k_{2})\big[(k_{1} - k_{2})^{\rho}g^{\mu\nu} + (k_{2} - k_{3})^{\mu}
g^{\nu\rho} + (k_{3} - k_{1})^{\nu}g^{\rho\mu}\big]\\
&- \theta^{\mu\nu}[k^{\rho}_{1}(k_{2}k_{3}) - k^{\rho}_{2}(k_{1}k_{3})]
 - \theta^{\nu\rho}[k^{\mu}_{2}(k_{3}k_{1}) - k^{\mu}_{3}(k_{2}k_{1})]\nonumber\\
&-  \theta^{\rho\mu}[k^{\nu}_{3}(k_{1}k_{2}) - k^{\nu}_{1}(k_{3}k_{2})]\nonumber\\
&+ (\theta k_{2})^{\mu}[g^{\nu\rho}k^{2}_{3} - k^{\nu}_{3}k^{\rho}_{3}]
 + (\theta k_{3})^{\mu}[g^{\nu\rho}k^{2}_{2} - k^{\nu}_{2}k^{\rho}_{2}]\nonumber\\
&+ (\theta k_{3})^{\nu}[g^{\mu\rho}k^{2}_{1} - k^{\mu}_{1}k^{\rho}_{1}]
 + (\theta k_{1})^{\nu}[g^{\mu\rho}k^{2}_{3} - k^{\mu}_{3}k^{\rho}_{3}]\nonumber\\
&+ (\theta k_{1})^{\rho}[g^{\mu\nu}k^{2}_{2} - k^{\mu}_{2}k^{\nu}_{2}]
 + (\theta k_{2})^{\rho}[g^{\mu\nu}k^{2}_{1} - k^{\mu}_{1}k^{\nu}_{1}]\nonumber\\
&+ \theta^{\mu\alpha}(ak_{1} + k_{2} + k_{3})_{\alpha}[g^{\nu\rho}(k_{3}k_{2})
 - k^{\nu}_{3}k^{\rho}_{2}]\\
&+ \theta^{\nu\alpha}(k_{1} + ak_{2} + k_{3})_{\alpha}[g^{\mu\rho}(k_{3}k_{1})
 - k^{\mu}_{3}k^{\rho}_{1}]\\
&+ \theta^{\rho\alpha}(k_{1} + k_{2} + ak_{3})_{\alpha}[g^{\mu\nu}(k_{2}k_{1})
 - k^{\mu}_{2}k^{\nu}_{1}],
\end{split}\end{equation}
and
\begin{equation}
\theta^{\mu\nu\rho} = \theta^{\mu\nu}\gamma^{\rho} + \theta^{\nu\rho}
\gamma^{\mu} + \theta^{\rho\mu}\gamma^{\nu}.\nonumber
\end{equation}
Energy momentum conservation dictates that unitarity has
to be satisfied~\cite{Ohl:2003zd}. The coupling of gauge boson to matter
field preserves the gauge condition in energy momentum 
conservation. Therefore, under the production of each
photon energy momentum $k^{\mu}_{1}$ and $k^{\nu}_{2}$ from the
above couplings, we obtain that energy momentum is
conserved when producing $k^{\rho}_{3}$ lagged momentum. Momentum
conservation in central mass frame is preserved on the coupling,
but the energy asymmetry is not conserved in electric field ambience.
The reason of the produced exotic energy is due to this coupling
proportional to the coupling constant multiplying the central energy.

Following the discussion, if the process conatins triple gauge
boson coupling by electric field. The exotic longitudinal state
in charged matter current is naturally produced from the
shifted charge ranged in its mass. The finial triple gauge boson
coupling stores sufficient exotic energy transferring from gauge
boson propagator to generate the non-physical state in the
finial gauge boson luminosity.

If we choose the central mass frame, the collider phenomenon 
does not be changed in this frame, even if $Lorentz$ invariance is
violated. We choose $\theta^{0i}$ = 0, and set observer standing on the
incident event. However, $k^{1}_{\mu}\theta^{\mu\nu}k^{2}_{\nu}$ is
useless in the devotion without $\theta^{0i}$.
Moreover, numerical section we introduce how the first order
$\theta_{\mu\nu}$ deformed term influences our results via
background magnetic field, $B^{i}$ = $\frac{1}{2}\epsilon^{ijk}\theta_{jk}$,
couples to photon polarization. The finial results of forward-backward
asymmetry is transparently indicated into parity violation effect.

\section{$e^+e^-$ $\to$ $\gamma\gamma$ physics}
We briefly review $e^{+}e^{-}$ $\to$ $\gamma\gamma$ process 
on noncommutative U(1) model\cite{Baek:2001ty}. The U(1) NCQED
is a complete order $\theta_{\mu\nu}$ deformed field theory with
containing even order $\theta_{\mu\nu}$ perturbation expansion.
However, we read that the event number is like a sinuous function with
parity is preserved in spite of containing triple photon coupling. It is
well-known that noncommutative geometry is a nonlocal perturbative
theory. It is seeming a phase transition in spacetime coordinates.
This dramatic phenomenon is a complete background deformed effect.

The unusual commutation relation induces a triple gauge boson coupling
on the electroweak model. Violating charge conservation, such as the
couplings $\gamma-\gamma-\gamma$ and $Z-\gamma-\gamma$, is
considered in amplitude\footnote{Because C($\gamma$) = C(Z) = -~1,
but preserves $CP$ symmetry}. U(1) gauge cannot produce parity violated
phenomenon without considering $Chern-Simons$ term in the Lagrange
or containing a non-equilibrium field in vacuum. There is no helicity violation
generated in this group if no parity violation effects taking
into account. The gauge field expansion are redefined as
\begin{equation}\label{eq12}
\hat A_{\mu} = A_{\mu} - \frac{1}{2}\theta^{\alpha\beta}A_{\alpha}
(\partial_{\beta}A_{\mu} + F_{\beta\mu}),
\end{equation}
and its strength field
\begin{equation}\label{eq13}
\hat F_{\mu\nu} = F_{\mu\nu} - \theta^{\alpha\beta}
(A_{\alpha}\partial_{\beta}F_{\mu\nu} + F_{\mu\alpha}F_{\beta\nu}),
\end{equation}
which it is a first order $\theta_{\mu\nu}$ expansion, where the background 
tensor is denoted by Eq.(1) and (2).

The polarization sum is revised to be the transition involved into noncommutative
phase,
\begin{equation}\label{eq14}
\sum_{s}\epsilon^{\star s}_{\mu}(k)\star\epsilon^{s}_{\nu}(k) = -\big{(}g_{\mu\nu} - \frac{n_{\mu}k_{\nu}
+n_{\nu}k_{\mu}}{n\cdot k} + \frac{n^{2}k_{\mu}k_{\nu}}{(n\cdot k)^{2}}\big{)},
\end{equation}
the noncommutative phase in front of the polarization sum gives us a lots
clues of the coherent effect between photon polarization, but the induced
$\theta_{\mu\nu}$ phase transition is useless on the collider process.
In fact, although the U(1) model does not contain parity violated source
without $Chern-Simons$, the odd order theta deformed term 
will deviate on the loop process, such as magnetic dipole 
moment and electric dipole moment~\cite{Riad:2000vy}.

Physically speaking, the background magnetic field induces a spin-magnetic 
effect, the term of charge violated coupling is simultaneously 
violating parity symmetry. Even in U(1) model, the perturbative 
expansion corrects all parity violated events on the odd order 
$\theta_{\mu\nu}$ deformation. The even order $\theta_{\mu\nu}$ 
deformations only contribute on the cross section magnitude.
Therefore, it easily discovers that parity violated 
phenomenon on electron annihilation process is justified
from the order of $s\times\Lambda^{-2}_{C}$ in series expansion,
\begin{equation}\label{eq15}
\frac{d\sigma}{dzd\phi} = \frac{\alpha^{2}}{4s}\bigg{[}\frac{u}{t} + \frac{t}{u}
 -4\frac{u^{2} + t^{2}}{s^{2}}\sin^{2}(\frac{k_{1}\theta k_{2}}{2})\bigg{]},
\end{equation}
which the last term is the same as the Compton process in exchanging 
$p_{2}$ and $k_{1}$, where $u$ = $(p_{1} - k_{2})^{2}$, 
$t$ = $(p_{1} - k_{1})^{2}$, and $s$ = $(p_{1} + p_{2})^{2}$. 
These processes are only contributed by the background electric 
field. It implies that the finial state photon does not interact 
with the background magnetic field, and its deviation is coming 
from the interaction between background electric field with 
electric charge.

We explore that electric field interaction with $e^{+}$ 
and $e^{-}$ on the opposite influence by multiplying a constant $b$
before of the imagine component in spinor vector polarization.
If we choose the electric field direction perpendicular to the
incoming incident, the event number is maximum distributioned.
Expectedly, $\alpha_{E}$ = 0 does not contain $\phi$ dependent
effect, because the preferred direction parallels to the incident
axis. On the noncommutative electroweak model, due to the unitarity
condition on the triple gauge boson coupling, we have to omit the
background electric field automatically. However, in the U(1) case,
the total cross section is proportional to the $\theta_{\mu\nu}$
second order term. If no preserced $\theta_{\mu\nu}$ odd order
term in the result, therefore, no symmetry properties can be found.

Nonetheless, in noncommutative electroweak model, the term
retains in the finial result. Intuitively, the process generates parity
asymmetry effect. Following the diagrams
\begin{figure}[htbp] 
   \centering
   \includegraphics[width=3.5in]{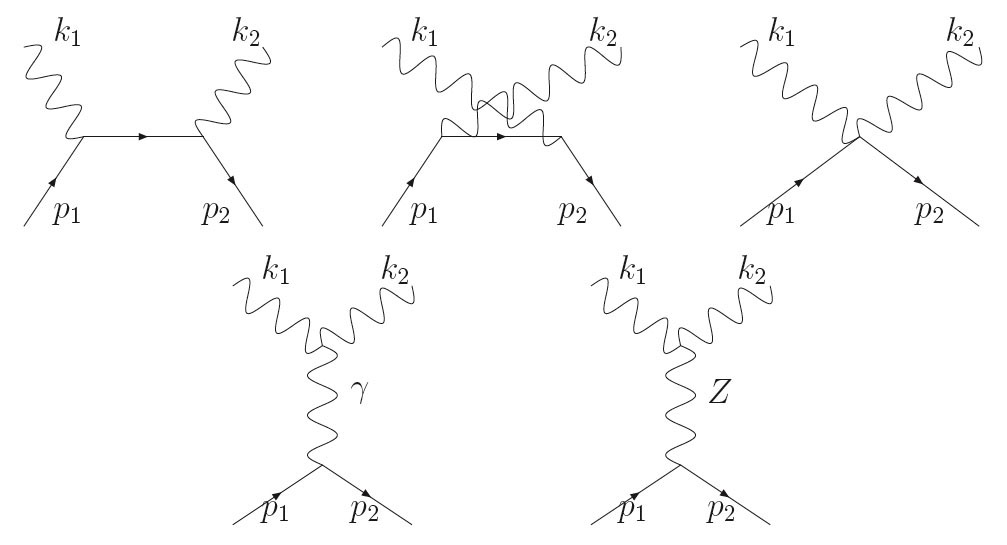} 
   \caption{The $e^{+}e^{-}\to\gamma\gamma$ diagrams}
   \label{fig:example}
\end{figure}
we write down the square amplitude and photon 
polarization under the first order $\theta_{\mu\nu}$ deformation.
Consider each photon polarization in
\begin{equation}\label{equ16}
\epsilon_{1 \mu} = (0, 1, b\textit{i}, 0),\quad\epsilon_{2 \mu}
= (0, 1, -b\textit{i}, 0),\\
\end{equation}
and each incoming momentum and outgoing momentum,
\begin{equation}\begin{split}
p^{\mu}_{1} = (E,0,0,E)&,\quad p^{\nu}_{2} = (E,0,0,-E),\\
k^{\mu}_{1} = E(1, \sin\theta&\cos\phi, \sin\theta\sin\phi, 1),\\
k^{\mu}_{2} = E(1,-\sin\theta&\cos\phi,-\sin\theta\sin\phi,-1),
\end{split}\end{equation}
in which $b$ is + or - corresponding to right-handed and left-handed
circle polarization with the incident working on the background
\begin{equation}
\vec{B} = \frac{1}{\Lambda_{C}}(\sin\alpha\sin\beta, \sin\alpha\sin
\beta, \cos\alpha).
\end{equation}
It is convenient to analyze the contribution of each different 
helicity.

We consider, $\theta^{ij}$, space-space noncommutative 
deformed spacetime. The total amplitude splites 
into a zero term and a first order theta deformation,
\begin{equation}
\sigma_{tot} = \sigma_{0} + \sigma_{\theta~}
\tt{(theta~first~order~term)},\nonumber%
\end{equation}
the first part is the original commutative term and the second is 
the first order theta deformed term. It contributes to the total 
cross section with a free gauge freedom constant "a" and helicity 
constant "b". The renormalization condition requires the 
parameter "a" to be 3. The cross section zeroth and first order 
term are as follows,
\begin{equation}\label{eq17}
\sigma_{0} = \frac{\alpha^{2}}{4s}\big(\frac{t}{u}+\frac{u}{t}\big),
\end{equation}
\begin{equation}\label{eq18}
\sigma_{\theta} = \frac{\alpha^{2}}{4s}\mathfrak{Re}\big[\sigma_{1}
+ (a+1)\sigma_{2}\big],
\end{equation}
where
\begin{equation}\label{eq19}
\sigma_{1} = -\frac{i}{2}(\epsilon_{1}\theta\epsilon_{2})\bigg{[}
\frac{s^{2}b\triangle}{2} + sz(s\square-1)\bigg{]},
\end{equation}
\begin{equation}\begin{split}\label{eq20}
\sigma_{2} &= \frac{i}{2}s^{\frac{3}{2}}\triangle\sqrt{\frac{1-z^{2}}{2}}\\
             &\bigg{[}\frac{(\epsilon_{1}\theta k_{1})(i\sin\phi - b\cos\phi)}
             {1+z}+\frac{(\epsilon_{2}\theta k_{1})(i\sin\phi+b\cos\phi)}{1-z}\bigg{]},
\end{split}\end{equation}
and
\begin{equation}\begin{split}\label{eq21}
\triangle &= \frac{2K_{Z\gamma\gamma}C_{A}}{s-m^{2}_{Z}},\\
\square  &= -\frac{2K_{\gamma\gamma\gamma}\sin2\theta_{W}}{s}-
\frac{2C_{V}K_{Z\gamma\gamma}}{s-m^{2}_{Z}},\\
C_{V, f_{L}} &= T_{3,f_{L}} - 2Q_{f}\sin^{2}\theta_{w},\nonumber\\
C_{A, f_{L}} &=T_{3,f_{L}}.
\end{split}\end{equation}
The total decay rate is contributed from 
complete theta second order modification. Hence, no asymmetry 
phenomenon can be generated. Under the rotation of the background field 
direction, the total decay rate is symmetrically rotated with the angular
momentum correlation between original axis and background unique direction. The 
$Z^{0}\to\gamma\gamma$ decay~\cite{Duplancic:2003hg} is completely
forbidden by angular conservation and bosonic distribution.

\section{Numerical Result}
In the numerical analysis, the influence of the background field direction
dominates the total cross section and differential cross section of each
helicity state. Each photon helicity interacts with the background
magnetic field in the opposite distribution. The asymmetry effects in the
finial helicity state are mutually canceled on the unpolarized cross
section. Moreover, concentrating on the result of the $\theta_{\mu\nu}$
deformed term, Eq.(16, 17, 18), we show that the $Z_{0}$ gauge boson mediator is
almost completely violating parity asymmetry. The contribution of the massless
gauge boson on each helicity state does not cause rapid changes. The cross
section is minutely varied, but, its behavior is dramatically changed on the
scattering process. Because the $Z^{0}$ gauge boson is working on a non-abelian
gauge and coupling to opposite helicity current by different distribution with the
mass approaching to 0.1 TeV.

\begin{figure}[htbp] 
   \centering
   \includegraphics[width=3in]{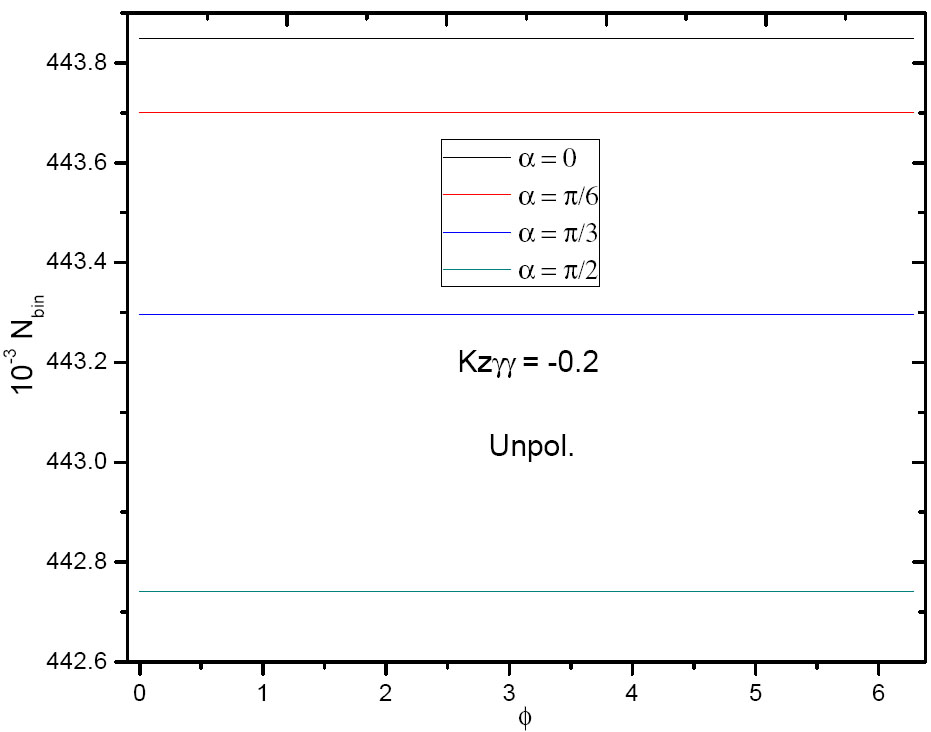} 
   \caption{The coupling constant $K_{Z\gamma\gamma}$ = -0.2, 
   where $E_{CE}$ = 800 GeV, $\Lambda_{C}$ = 1TeV. As the result 
   from $Z^{0}$ gauge boson couples to the matter current by 
   the different contribution, therefore it will contribute on the 
   unpolarized cross section. If $K_{Z\gamma\gamma}$ = 0 
   the number of event approaches to QED result 442.74.}
   \label{fig:example}
\end{figure}

In Fig.(2), the contribution of the $Z^{0}$ gauge boson 
process dominates the total cross section, and the results compare 
with the unpolarized beam in setting $K_{Z\gamma\gamma}$ = - 0.2. 
The $Z^{0}$ gauge mediator produce a slight shift, but the photon
sector will not be changed. In the $SU_{L}(2)\otimes U_{Y}(1)$ model, 
photon is a gauge boson coupling to each helicity current by the same 
phenomenon. $Z^{0}$ gauge boson induces a different distribution 
in the left-handed and right-handed currents. Therefore, on the
Left-Right symmetry model, $SU_{C}(3)\otimes SU_{R}(2)\otimes
SU_{L}(2)\otimes U_{Y}(1)$, the asymmetry effect wishes to disappear
on the unpolarized $Z^{0}$ channel. Due to unitary constraint on the gauge
sector, it should be conserved on the SU(N) group. Hence, Seiberg-Witten
map cannot give the other clues to allure us to do the work in extending
gauge sector from choosing larger $\theta_{\mu\nu}$ expansion.

\begin{figure}[htbp] 
   \centering
   \includegraphics[width=3in]{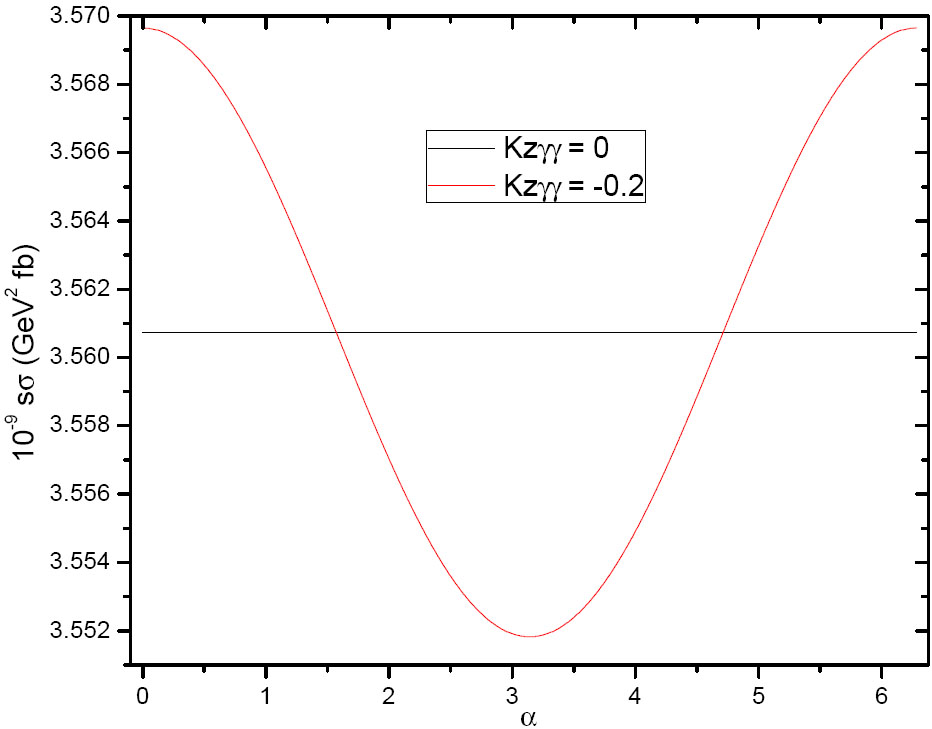} 
   \caption{The coupling constant $K_{Z\gamma\gamma}$ = 0 and -0.2, 
   and the central energy $E_{CM}$ = 800 GeV at $\Lambda_{C}$ = 1TeV
   scale. As $K_{Z\gamma\gamma}$ = 0, that will be as same as QED result, 
   3.561 unit. In the polarized helicial state, the contribution of b = 1, and 
   b = -1 on the background field direction along the z-axis are the same.}
   \label{fig:example}
\end{figure}
\begin{figure}[htbp] 
   \centering
   \includegraphics[width=3in]{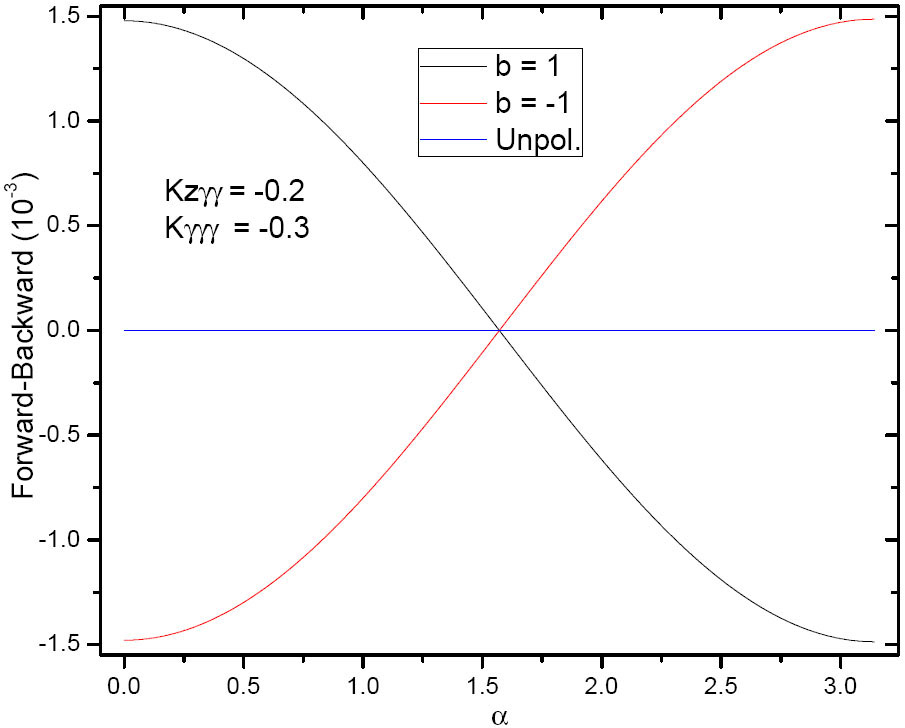} 
   \caption{The forward-backward asymmetry is mainly affected 
   by the cube vertex diagram, the $Z^{0}$ mediator gauge boson 
   contributed effect is actually very small. The coupling constants 
   $K_{Z\gamma\gamma}$ = -0.2, $K_{\gamma\gamma\gamma}$ = -0.3, 
   and cnetral energy $E_{CM}$ = 800 GeV at $\Lambda_{C}$ = 1TeV.}
   \label{fig:example}
\end{figure}

The total cross section cannot be corrected with setting 
$K_{\gamma\gamma\gamma}$ = 0 in Fig.(3), since photon 
is a complete U(1) gauge boson. In high energy level, 
the main distribution presents along the z-axis. As to the $\phi$-axis, 
the influence of the spin-magnetic interaction is very little influenced. 
Visibly, the diagram, Fig.(3), is a perfect symmetry on the 
limit point $\alpha$ = $\pi$. It is a result in assuming two 
observers stand on the either sides of the event point. 
They cannot get the same result as detecting the total cross 
section of each photon helicity. The order of difference quality is associated 
with the squared inverse of the $\Lambda_{c}$ parameter. 
Throughout the F-B asymmetry discussion, we set the parameter 
$\Lambda_{c}$ to 1 TeV, and the central energy is assumed to be 800 GeV.

\begin{figure}[htbp] 
   \centering
   \includegraphics[width=3in]{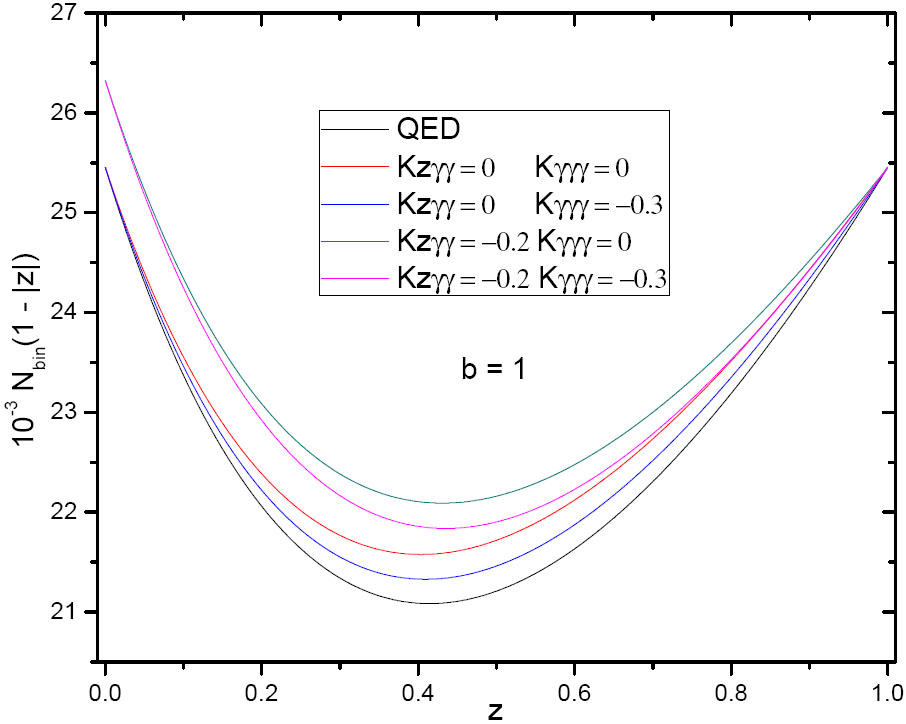} 
   \caption{The coupling constant $K_{Z\gamma\gamma}$ = -0.2 and 0, the 
   $\alpha$ = $\frac{\pi}{3}$ and $E_{CE}$ = 800GeV at $\Lambda_{C}$ = 
   1TeV. The black dashline is the original QED prediction.}
   \label{fig:example}
\end{figure}
\begin{figure}[htbp] 
   \centering
   \includegraphics[width=3in]{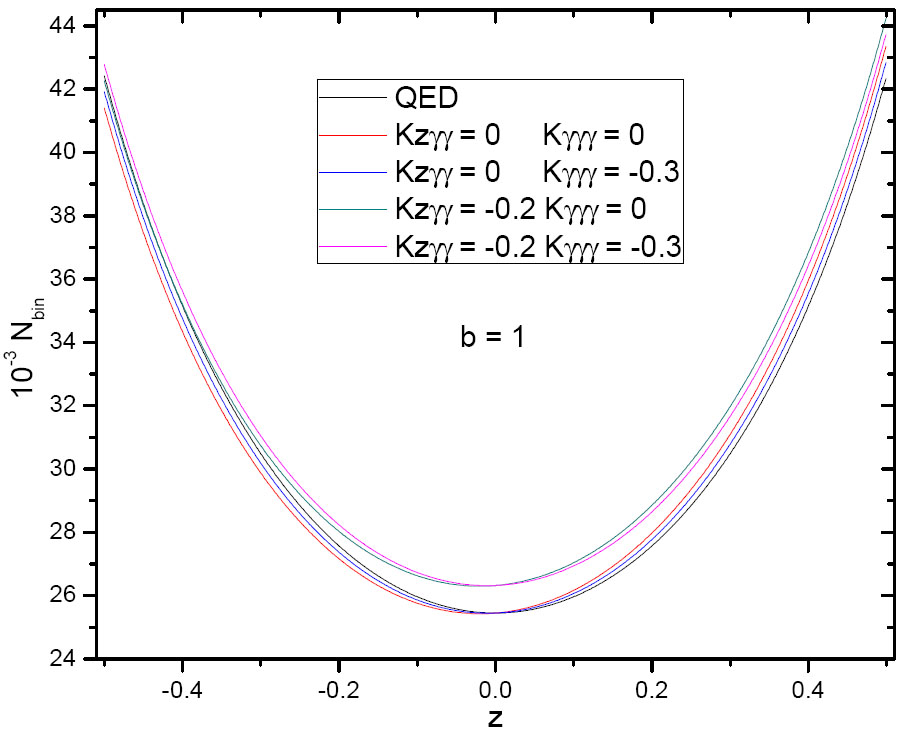} 
   \caption{The coupling $K_{Z\gamma\gamma}$ = -0.2, 0 and 
    $K_{\gamma\gamma\gamma}$ = -0.3, 0, where $\alpha$ =
    $\frac{\pi}{3}$, $E_{CE}$ = 800GeV, and $\Lambda_{C}$ = 1TeV. 
    Comparing to the QED result, the coupling $K_{\gamma\gamma\gamma}$ 
    will be dominant in influency the slight shift effects.}
   \label{fig:example}
\end{figure}
    
The main idea of parity violation, Fig.(4)(5), is a spin-magnetic field
interaction. Which is contributed on the difference energy
distributions on the opposite sides of the event point. If spin orientation
is parallel to the background magnetic field, the energy distribution is maximally
contributed. In contrast, energy is diminished if the direction
between spin and background magnetic field is opposed with the angle depends
on the z-axis and the background preference. It is the reason
why we can get parity violation phenomena. The term of $\vec{S}\times\vec{B}$ 
gives us a different physics viewpoint to investigate the process. 
This term clearly indicates spin cannot be perpendicular to 
the background magnetic field. The difference of varying parity asymmetry
associates with the strength of background magnetic field.

\begin{figure}[htbp] 
   \centering
   \includegraphics[width=3in]{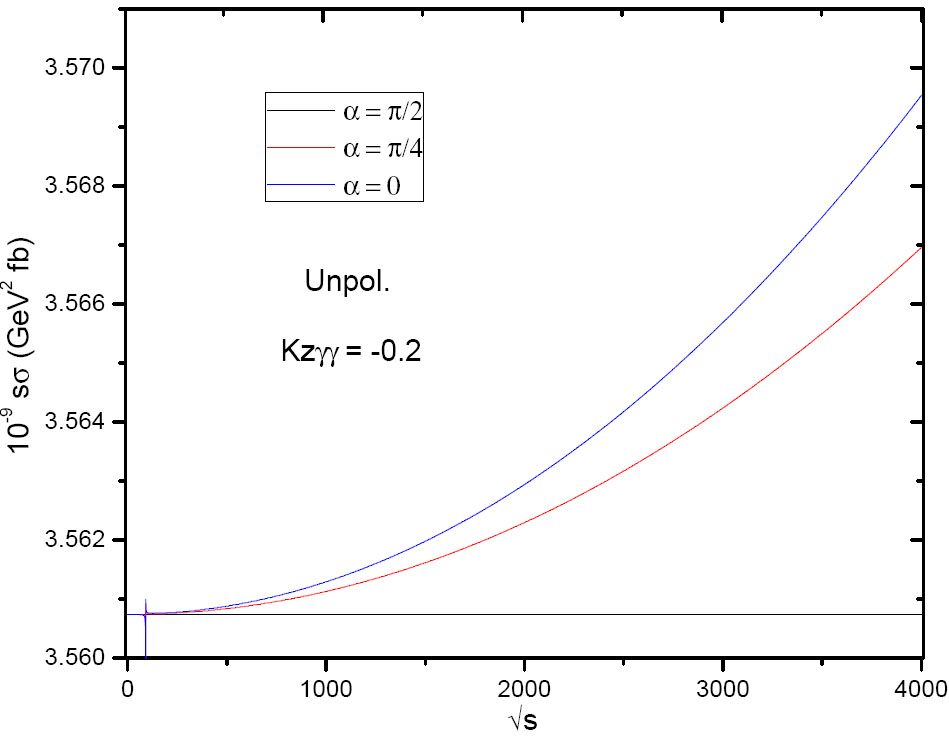} 
   \caption{The Unpolarized cross section, we set the coupling constant 
   $K_{Z\gamma\gamma}$ = -0.2, where the $\Lambda_{C}$ is set to 
   5000 GeV. The energy spectrum is increasd by the spin-magnetic 
   production effect, gauge boson polarization coupling to the background 
   magnetic field on the event point will rearrange the distribution of 
   energy production.}
   \label{fig:example}
\end{figure}

Another, observable evidences, Fig.(6), are the quantity of event number 
as to the z variable, z = $\cos\theta$.  The helicity is contributed by
the parity asymmetry effects on finial result. We consider helicity state
$b$ = 1 and discuss, however, that another helicity state b = -1 is
shifted on the opposite side. If we set the $K_{\gamma\gamma\gamma}$
coupling equals to zero then the signal is similarly unchanged, because
$Z^{0}$ boson is heavier than photon. Photon gauge boson contributes to the
finial result has perceived more than massive $Z^{0}$. The shift devotes
on the photon spin interacts with background magnetic field, and the 
axis is perpendicular to the direction. The external distribution is perpendicular
to the axis of the magnetic field direction, because the effective term of
$\vec{S}\times\vec{B}$ generates a partial vector paralleling to the plane.

The energy spectrum, Fig.(7), in ranging the central energy, the
$\theta_{\mu\nu}$ expansion plays an important role on varying
the associated angle. The $\alpha$ = 0 generates a distribution at high energy
level because the polarization of the total cross section on the event point
is parallel to the beam axis. However, we have mentioned that if the
polarization is parallel to the magnetic field, thus, the result obtains the maximum
energy distribution function. Such as the concept of quantum mechanics,
the energy spectrum is decided by the eigenvalues of the global system.
Therefore, $\alpha$ = 0, on the event point, photon gain a maximum
energy distribution on the collision process. Its luminosity contains tiny
difference as to the movement of earth.

\section{Conclusion}
We have briefly introduced how the background magnetic field influences 
the electron annihilation to two photons process. A strong magnetic field
induces an interesting effect under the exotic massive gauge boson $Z^{0}$
and massless photon. However, parity violation is observed on the
further high energy level. $CP$ symmetry is still conserved on the triple
photon and $Z^{0}$ gauge boson coupling, due to these couplings violate
charge and parity asymmetry. Thus, the exotic term in the action deformed
by $\theta_{\mu\nu}$ expansion cannot induce the $CPV$ effects. However
, the energy spectrum, due to particle spin, interacts with magnetic field
to generate a difference energy distribution on the opposite sides around the
event point. The energy distribution dominantly induces the parity asymmetry
on the observer stage. Therefore, this process is a contribution of a better
understanding of further probing background field situation.

\acknowledgments{We will thank Chao Qiang Geng, Xiao Gang He, and J. N.
Ng for useful discuss and the National Science Council of R.O.C. under
contact : NSC-95-2112-M-007-059-MY3 and National Tsing Hua University
under contact : 97N2309F1.}

\end{document}